\documentclass[twoside,twocolumn,english, floatfix, longbibliography, superscriptaddress, prb, 10pt]{revtex4-2}
\usepackage[T1]{fontenc}
\usepackage{color}
\usepackage{graphicx}
\usepackage{babel}
\usepackage{amsbsy, amsmath, amssymb}

\begin{document}
\title{Spin-wave localization on phasonic defects\\ in one-dimensional magnonic quasicrystal }
\begin{abstract}
We report on the evolution of the spin-wave spectrum under structural disorder introduced intentionally into one-dimensional magnonic quasicrystal. 
We study theoretically a system composed of ferromagnetic strips arranged in a Fibonacci sequence. We considered several stages of disorder in the form of phasonic defects, where different rearrangements of strips are introduced.
By transition from the quasiperiodic order towards disorder, we show a gradual degradation of spin-waves fractal spectra and closing of the frequency gaps. In particular, the phasonic defects lead to the disappearance of the van Hove singularities at the frequency gap edges by moving modes into the frequency gaps and appearing new modes inside the frequency gaps. These modes disperse and eventually can close the gap, with increasing disorder levels. The work reveals how the the presence of disorder modifies the intrinsic spin wave localization existing in undefected magnonic quasicrystals. The paper contributes to the knowledge of magnonic Fibonacci quasicrystals and opens the way to study of the phasonic defects in two-dimensional magnonic quasicrystals.

\end{abstract}
\author{Szymon Mieszczak}
\email{szymon.mieszczak@amu.edu.pl}
\author{Maciej Krawczyk}
\author{Jaros\l aw W. K\l os}

\affiliation{Institute of Spintronics and Quantum Information, Faculty of Physics, Adam Mickiewicz University, Pozna\'{n}, Uniwersytetu Pozna\'{n}skiego 2, Pozna\'{n} 61-614, Poland}
\maketitle

\section{Introduction}
Quasicrystals are aperiodic structures characterized by long-range order and lack of translational symmetry\cite{Shechtman1984, Levine1984}. The order can be revealed in the Fourier spectrum of the structure that has a countable set of Fourier components\cite{Janot2012, Vardeny2013, Negro2014, Janssen2007}. 
This property leads to the presence of multiple frequency gaps (i.e., Bragg gaps) in the spectrum of eigenmodes.
The disorder introduced into the structure generally leads to the localization of the eigenmodes. The increasing level of disorder eventually leads to Anderson localization\cite{Anderson1958, Levi2011, Segev2013, Kromer2012} and the gradual closing of the Bragg gaps. Particularly interesting is the case of defects in quasicrystals because they possess fine band structures and already localized modes that are called critically localized. In this sense, the impact of the disorder can be more complex. 

 Due to the structural degrees of freedom in quasicrystals, the local arrangement of the structure cannot unambiguously determine the global ordering and the identification of disorder is more difficult than for periodic structures. The concept of structural degrees of freedom is more understandable when we notice that the quasicrystals can be generated from the higher-dimensional crystals defined in abstract higher-dimensional hyperspace or real space but by the cut-and-projection (C\&P) method\cite{Janot2012}. 
 
 The most known 1D quasicrystal whose lattice can be generated by the C\&P method is the Fibonacci quasicrystal, where lattice points, separated by long ($L=a\tau/\sqrt{2+\tau}$) and short ($S=a/\sqrt{2+\tau}$) distances, are arranged aperiodically ($a$ denotes the period of square lattice in hyperspace, $\tau$ is the golden ratio)\cite{Jagannathan2021}. The translation of the Fibonacci lattice is equivalent to rearrangements/swaps within the pairs of neighboring sites, which leads to the exchange of the adjacent short and long distances: $LS\leftrightarrow SL$. These local rearrangements of the lattice are called {\em phasons}\cite{Maci__2005}. 
 The C\&P method suggests also how to generate the positional disorder in the Fibonacci lattice manifested only by the perturbation of the sequence of $L$ and $S$. It can be achieved by the modulation of the shift $c$ of the projection line $y=\tau^{-1}x+c(x)$ -- see Appendix~\ref{sec:app-C-P} for more details. If this randomly introduced modulation is long-wave and has small amplitude, then it generates the $LS\leftrightarrow SL$ swaps. Such kind of structural disorder is called {\em phasonic defects}.
  
The phasons (and phasonic defects) are the unique feature of all quasicrystals and were intensively investigated in relation to the stability of atomic lattice of natural quasicrystals and their phononic properties\cite{Socolar1986}. In these systems, phasons are dynamic objects which can be activated thermally and move diffusely\cite{Boissieu2012, Kromer2012, Wolny2016} in the structure of quasicrystal.
The concept of phasons was already investigated in photonics including the diffusive character of phasons\cite{freedman_phason_2007}. Their role was also discussed as static defects, deliberately introduced into the photonic quasicrystals\cite{Bandres2016}. 

In the paper, we focus on the general problem of proper introduction of positional disorder in magnonic quasicrystals and study the impact of such phasonic defects on the spin-wave spectra and their localization properties in magnonic Fibonacci quasicrystals\cite{Chen2014, Rychly2015}.
We introduce the static and spatially uncorrelated phasonic defects, which allows for gradual transition from non-defected Fibonacci sequence of strips to the completely disordered system. The static character of considered phasonic defects means that they are introduced intentionally (i.e. by design) and not spontaneously (i.e. by thermal activation).

The impact of the disorder on magnetization dynamics was extensively studied in the lattice models\cite{Ding2011, Evers15, Evers2018, Buczek_2018}.
In the case of the continuous model, the impact of the isolated defect on the spin-wave spectrum in magnonic crystals was investigated for 1D structures \cite{Gallardo2018, Tkachenko2010, Kruglyak06}, 2D magnonic crystals\cite{Yang12a, Yang12b, YANG14}, and line defects in 2D magnonic crystals\cite{XING15}. There were also reports on defect as a magnetization reversal of a single strip in one-dimensional magnonic crystal\cite{Baumgaertl18}. However, a disorder in magnonic quasicrystals raises another class of questions thus, we believe that our study on phasonic defects and their impact on the spin-waves makes a valuable contribution to the magnonics field of research.

In the section 'Structure and Model', we present the magnonic structure
under investigation, explaining (i) why this structure can be considered as a decorated Fibonacci lattice and (ii) how we introduce the uncorrelated phasonic defects. In this section, we also outline the computational method based on the solution of the Landau-Lifshitz equation by the plane wave methods. In the section 'Results and Discussion', we provide a detailed analysis of the impact of the phasonic defect on the frequency spectra of SWs and localization of the modes, illustrated by the plots of the integrated density of states, localization measure, and the profiles of selected modes. Finally, in the last section 'Summary', we conclude our findings.

\section{Structure and model}
We investigate spin-waves (SWs) in a 1D planar
magnonic structure composed of cobalt (Co) and permalloy (Py) strips of equal widths, being in direct contact and thus forming a continuous layer. The Co and Py strips are magnetically saturated by the external field applied along with them. The strips are arranged in Fibonacci quasicrystal. 
It is worth noting that despite the equal width of the strips, the system can be understood as a decorated Fibonacci lattice where Co and Py strips are centered at sequences
$SLLS$ and $SLS$ 
($S=a/\sqrt{2+\tau}$, $L=a\tau/\sqrt{2+\tau}$, where $a$ is the period in hyperspace) sharing the shorter sections $S$ between Co and Py with the ratio: $(2-\tau)/(2+\tau)$. 
Then, the common width of Co and Py strips is equal to $a(\frac{3}{2}\tau+1)/\sqrt{\tau+2}$.

\begin{figure}[t]
\begin{centering}
\includegraphics[width=0.9\columnwidth]{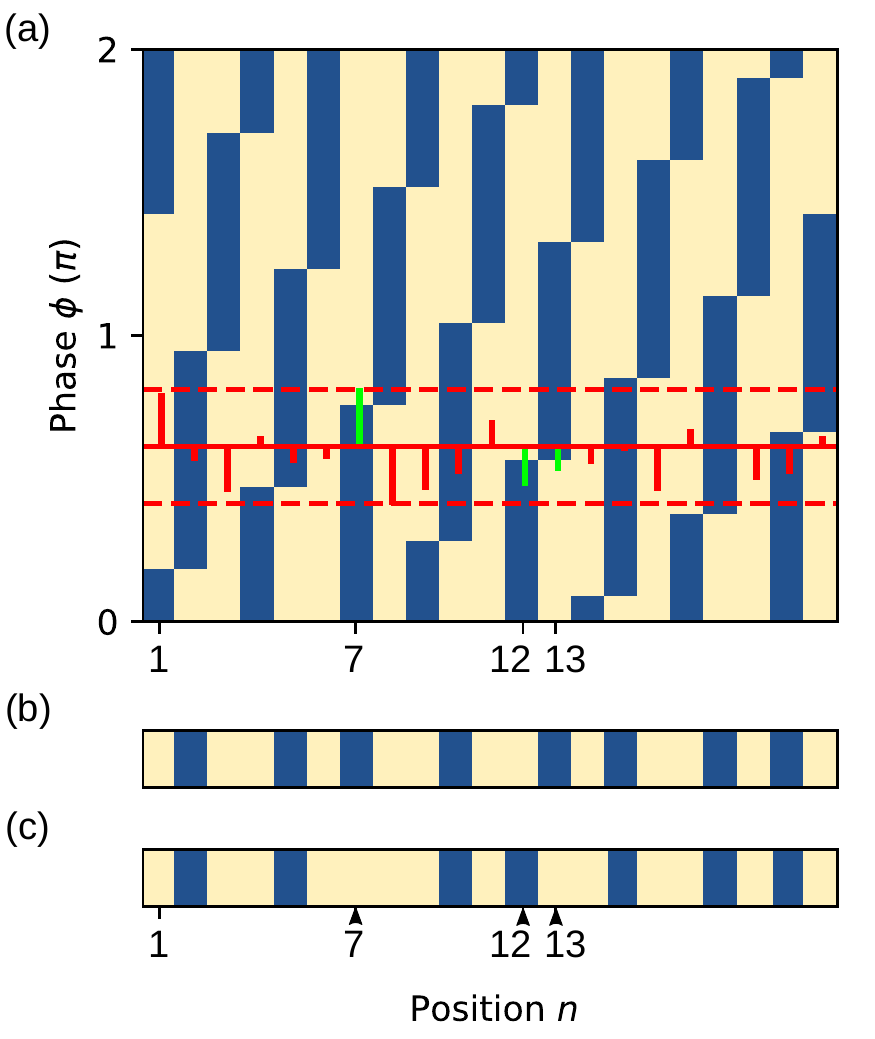}\caption{\label{fig:phasons_introduction}
(a) The all possible approximates of Fibonacci crystal composed of 21 elements. As the phase $\phi$ changes (see Eq.~(\ref{eq:chi})), we obtain 21 possible sequences of Co (light yellow) and Py (dark blue) -- note that Co strips can appear in doublets. The solid red line at $\phi=\pi/\tau$ corresponds to the approximate generated by standard substitution rules: ${\rm Co}\rightarrow{\rm Co}|{\rm Py}$, ${\rm Py}\rightarrow{\rm Co}$, presented in (b) -- see also Fig.~\ref{fig:fib21}. The red dashed lines show the range in which the parameter $\phi$ is randomly changed at each position $n$. The changes of $\phi$ which induces the phasonic defects are marked by green bars. They are responsible for the substitution ${\rm Py}\rightarrow {\rm Co}$ at position 7 and swap between positions 12 and 13 (${\rm Co|Py}\rightarrow {\rm Py|Co}$).
Sequence with defects is presented in (c); note that position of swaps are marked by arrows.}  
\par\end{centering}
\end{figure}

\begin{figure}[t]
\begin{centering}
\includegraphics[width=1\columnwidth]{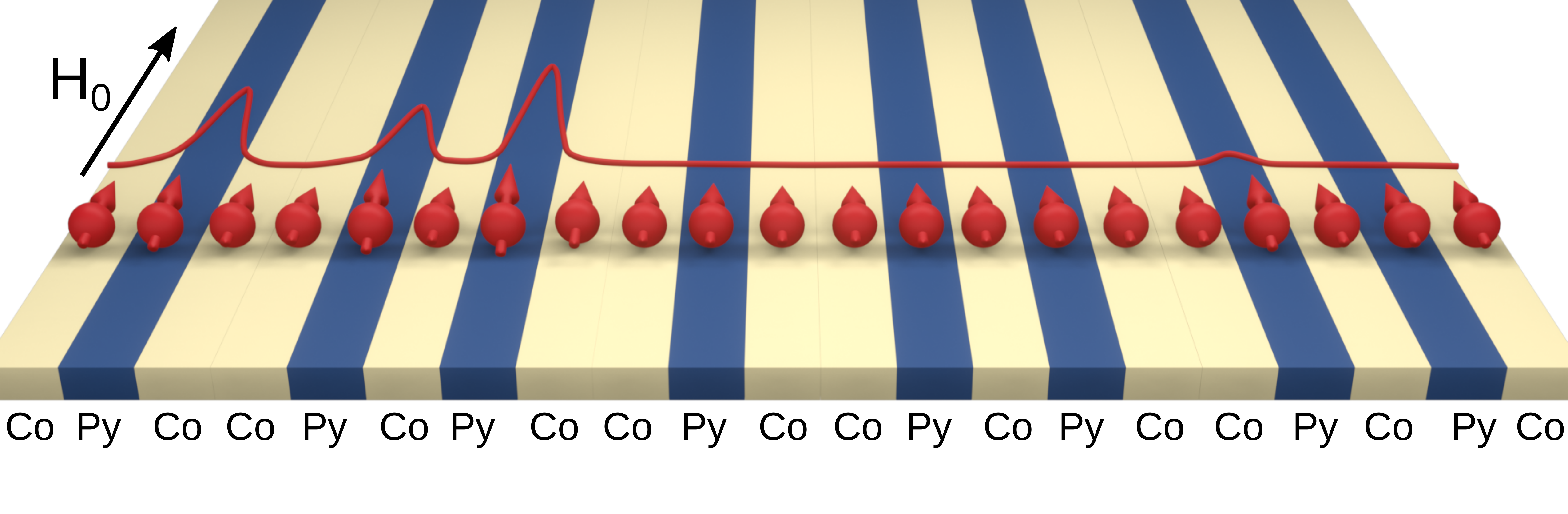}\caption{\label{fig:fib21}
The approximate of Fibonacci quasicrystal corresponding to the $\phi=\pi/\tau$ (see Fig.~\ref{fig:phasons_introduction}), i.e., resulting from the standard substitution rules: ${\rm Co}\rightarrow{\rm Co}|{\rm Py}$, ${\rm Py}\rightarrow{\rm Co}$. This exemplary structure is composed of Py and Co flat strips (30~nm thick and 300~nm wide), aligned side-by-side and being in direct contact. The field $\mu_0H_0$=0.1 T is applied along the strips. The sequence of tilted arrows and line in front of them visualizes the spin-wave mode profile. 
}
\par\end{centering}
\end{figure}

To generate the phasonic defects as the structural perturbations, we use the procedure which is technically simpler than the C\&P method (discussed in Appendix A), although it is based on more complex formalism (describing the properties of the generalized Harper model with incommensurate modulation of the on-diagonal and off-diagonal elements of tight-binding  Hamiltonian\cite{Kraus2012a}). The general model, which also describes Fibonacci quasicrystal, is presented in Ref.~\onlinecite{Kraus2012a}. 
The authors provide the characteristic equation that determines the successive elements of the Fibonacci sequence for given values of the parameter $\phi$, describing the structural degree of freedom\cite{Dareau17, levy_topological_2016}:
\begin{equation}
    \chi_n (\phi) = \text{sign} \left[ \cos \left( \frac{2 \pi n}{\tau} + \phi \right) 
    - \cos \left(\frac{\pi}{\tau} \right)\right].\label{eq:chi}
\end{equation}
The characteristic function $\chi_n$ takes the values $\pm 1$.  For our structure, $\chi_n=1$ ($\chi_n=-1$) selects Py (Co) strip at $n{\rm^{th}}$ position  in the Fibonacci sequence. 
 The  parameter $\phi$ is related to the shift $c$ of the line $y=\tau^{-1}x+c$ in C\&P method: $\phi=2\pi c/a$, see Fig.~\ref{fig: CaP} in Appendix A. 
 For infinite range of the index $n$, the different values of $\phi$ correspond to different realizations of the Fibonacci crystals which are only shifted by $\tilde{n}$ positions with respect to each other: $\chi_n(\phi)=\chi_{n+\tilde{n}}(\phi+\tilde{\phi})$, for every $n$ (the change $\tilde{\phi}$ of the parameter $\phi$ corresponds to the shift of the sequence by $\tilde{n}$ positions).
 When $n$ takes values in the finite range $1,\ldots,N$, and $N$ is the Fibonacci number,
 then the sweep of the parameter $\phi$ in the range $[0,2\pi]$ produces all unique, $N$-element sequences which can be identified as the Fibonacci crystal.
 The number of such unique approximates of Fibonacci crystal is equal to $N$. It is illustrated in Fig.~\ref{fig:phasons_introduction}(a) where we presented all 21 approximates composed of 21 elements (strips). Please note that the  characteristic function  Eq.~(\ref{eq:chi}) is periodic: $\chi_n(\phi)=\chi_n(\phi+2\pi)$, and the parameter $\phi$ plays a role of phase in Eq.~(\ref{eq:chi}).

 We arbitrarily selected the structure represented by $\phi=\pi/\tau$ because this approximate is generated by the standard substitution rules. The phasonic defects can be introduced to any sequence generated by Eq.~(\ref{eq:chi}), because each of them is defect-less section of Fibonacci quasicrystal. The approximate for $\phi=\pi/\tau$ (red solid line in Fig.~\ref{fig:phasons_introduction}(a)) is presented schematically in Fig.~\ref{fig:phasons_introduction}(b) and corresponding structure is visualized in Fig.~\ref{fig:fib21}.

To introduce the phasonic defects we add an additional term $\phi_n$ to the parameter $\phi$: $\phi\rightarrow~\phi+\tilde{\phi}+\phi_n$\cite{Dareau17,levy_topological_2016}. 
This term is a random number of uniform distribution in the range  $-\Delta\phi<\phi_n<\Delta\phi$, where $\Delta\phi<\pi$.  The range $\Delta \phi$ can be understood as a counterpart of thermodynamic temperature in atomistic quasicrystals, where higher temperature lead to higher probability of defect occurrence.
This range is marked by the red dashed lines in Fig.~\ref{fig:phasons_introduction}(a) and the exemplary sample of the random values of $\phi_n$ are denoted by thin vertical bars. 
The perturbations $\phi_n$ which induce the phasonic defects (i.e., flip the sign of $\chi_n$) are marked by the green line (positions $n=7,12,13$). 
The ineffective perturbations are marked by red bars. The perturbed structures with three phasonic defects are shown in Fig.~\ref{fig:phasons_introduction}(c). Position of phasonic defects are marked by arrows below the figure. 
The defects are not correlated in space because for each position $\chi_n$ is generated independently. 
Thus, the parameter $\phi_n$ does not change gradually, in a wave-like manner, as it is expected for a long wave (and long-living) phasons in atomic quasicrystals at finite temperature\cite{Boissieu2012, Socolar1986}. Because of it, along with swaps $LS\leftrightarrow SL$, we can also observe the substitutions $L\leftrightarrow S$.
For $\Delta \phi = \pi$ the system becomes random, since the probability of type of strip at $n-$th position is $\tau$. We discuss this case in Appendix~B. 
For smaller values of the amplitude $\Delta \phi$, the introduction of defect is not equally probable at every position. At some locations (e.g., the position 13 in Fig.~\ref{fig:phasons_introduction}) the generation of defect is highly probable whereas other locations can be quite robust (e.g., position 7), or even completely inaccessible (e.g., position 2) for defects\cite{Naumis2007}. 

Each strip is assumed to have a width of $300$~nm, a thickness of $30$~nm, and is infinitely long. The dimensions make the system in an exchange-dipolar regime, which is already feasible for experimental realization. For the constituent elements from which the system is constructed, we consider two widely used materials, namely Co and Py. The parameters that are important
for SW propagation are magnetization saturation  $M_{\rm S}$ and the exchange length $\lambda_{\rm ex}$. These parameters are equal to
$M_{\rm S,Co}=1445$~kA/m, $\lambda_{\rm ex,Co}=4.78$~nm, $M_{\rm S,Py}=860$~kA/m, and $\lambda_{\rm ex,Py}=5.29$~nm. We assume that our sample is saturated
by the external magnetic field with value $\mu_{0}H_{0}=0.1$~T, and is directed along the strips. In this geometry, a static demagnetizing field is equal to zero.

We consider magnonic quasicrystal that is composed of two different magnetic materials\cite{Wang09, Choudhury16}. However, the magnetic contrast can also be obtained in other ways: by inducing locally anisotropy\cite{Wawro18, Frackawiak20}, by decorating the uniform film\cite{Graczyk18, Liu18} or by thermal gradient\cite{Chang18}.
Having said that, the physics that we present in the paper is not restricted to the bi-component material. 

When we neglect the damping, the dynamics of magnetization vector can be described by Landau-Lifshitz
equation (LLE):
\begin{equation}
\frac{\partial\boldsymbol{M}}{\partial t}=-\mu_{0}\left|\gamma\right|\boldsymbol{M}\times\boldsymbol{H}_{\rm eff},\label{eq:LLE}
\end{equation}
where $\mu_{0}=4\pi\times10^{-7}$~H/m is the permeability of vacuum
and $\gamma=176$~rad GHz/T is the giromagnetic ratio. The effective magnetic field, which
contains all kinds of magnetic interactions considered in our study,
governs the precession of magnetization vector.
In our case $\boldsymbol{H}_{\rm eff}$ is composed of the following terms:
\begin{equation}
\boldsymbol{H}_{\rm eff}\left(\boldsymbol{r},t\right)=\boldsymbol{H}_{0}+\boldsymbol{H}_{\rm dm}\left(\boldsymbol{r},t\right)+\boldsymbol{H}_{\rm ex}\left(\boldsymbol{r},t\right),
\end{equation}
where $\boldsymbol{H}_{0}$ stands for the external field, $\boldsymbol{H}_{\rm dm}\left(\boldsymbol{r},t\right)$
demagnetizing field and $\boldsymbol{H}_{\rm ex}\left(\boldsymbol{r},t\right)$
is the exchange field. The last two terms are spatially and temporally dependent
since they are connected with material parameters and magnetization dynamics at
the same time. 
SWs are usually studied at room temperatures $T$. Considered materials have much higher Curie temperatures $T_C$, e.g., $T_C\approx$1400 K for Cobalt. In the regime $T\lesssim3/4T_C$, thermal effects can be neglected, and the usage of Landau-Lifshitz equation is fully justified\cite{Evans12}.

We use the plane wave method (PWM) to solve the linearized LLE\cite{Krawczyk2012}, where the magnetization vector $\boldsymbol{M}(\boldsymbol{r},t)$ can be decomposed to static part $M_0(\boldsymbol{r})$ and dynamic $\boldsymbol{m}(\boldsymbol{r})e^{i 2\pi f t}$, changing harmonically with the frequency $f$.
Dynamic part contains two components of magnetization vector: $m_{int}(\boldsymbol{r},t)$, and $m_{out}(\boldsymbol{r},t)$, representing in-plane and out-of-plane oscillation, respectively.
The PWM method is designed for a periodic system, where the Bloch boundary condition must be used. The PWM is based on the application of the Fourier transform both to the Bloch functions (dynamic components of magnetization) and material parameters (saturation magnetization and exchange length). These procedure allows to formulate an algebraic eigenproblem which can be solved numerically with the eignevalues (being eigenfrequencies) and eigenvectors (being the Fourier coefficients of the Bloch functions).

Despite the fact that the quasicrystals are not periodic structures, the PWM can still be used in the so-called supercell approach\cite{Klos11}. This application of PWM still assumes periodicity, but for supercells being a copies of the whole system, for which we take the periodic boundary condition. 
In magnonics, this approach was already used to investigate defect modes\cite{Gallardo18}, interface modes\cite{Mieszczak2022}, waveguides\cite{Pan2017}, and two-dimensional quasicrystals\cite{RYCHLY2018, Watanabe20}.
For considered system, the supercells are composed of 377 strips. For such large supercells the peculiarities of the Fibonacci quasicrystal are well reproduced, and spurious interface states (which can appear at the edges of supercells) do not disturb the spectra.
We used 3770 plane waves for expansion into the Fourier series. This amount was checked for convergence and was enough to reproduce Fibonacci spectra\cite{Rychly2015}.

\section{Results and discussion}

\begin{figure*}
\begin{centering}
\includegraphics[width=0.8\paperwidth]{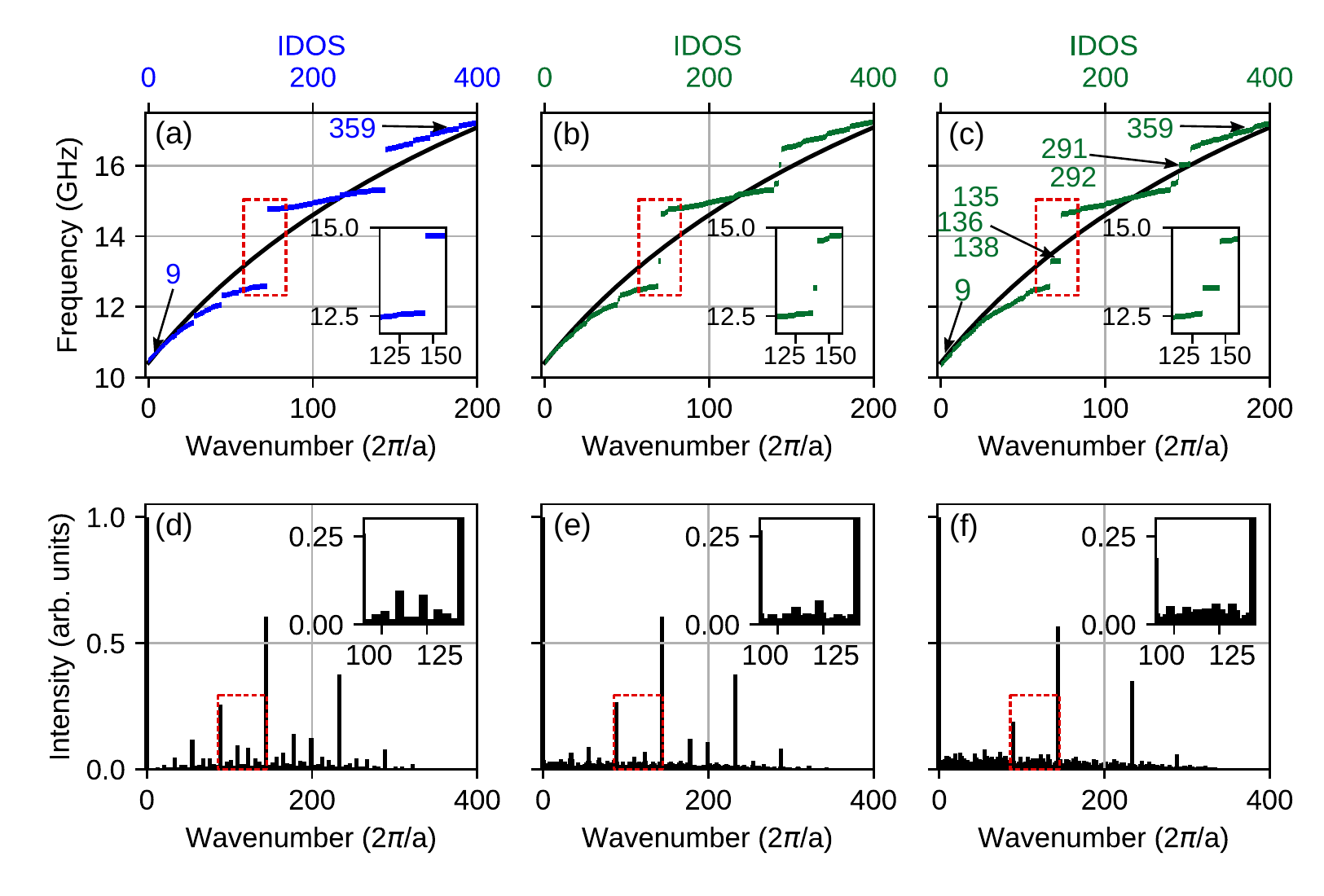}\caption{\label{fig:bragg}
Top row: (a), (b), and (c)  -- integrated density of states as a function of frequency plotted in the inverse form $f(\rm IDOS)$ -- blue/green color, and the dispersion relation for SWs in a homogeneous film with weight averaged material parameters -- black color. 
Please note that each information has its own abscissa, and share common ordinate.
The abscissa of each plot has the same scale, indicated only in the leftmost plot.
(a) Results obtained for perfect Fibonacci sequence composed of 377 strips.
(b) Results obtained for defected sequence with amplitude $\Delta\phi/(2\pi)=$ 5\%, and
(c) $\Delta\phi/(2\pi)=$ 10\%.
Bottom row: (d), (e), and (f) -- bar plot of reciprocal lattice vectors intensities, corresponding to the Bragg peaks, for the structures from (a), (b), and (c), respectively.
Phasonic defects destroys the fine structure of Bragg peaks that in consequence lead to modification of the density of states at the edges of frequency gaps and appearing new modes inside the gaps. 
}
\par\end{centering}
\end{figure*}
To determine the spectral properties of the approximates of Fibonacci quasicrystal, we plotted the dependence of integrated density of states (IDOS) on the frequency. For a finite system, ${\rm IDOS}(f)$ is the number of the modes below given frequency $f$ -- see Refs.~\onlinecite{Rychly2015, RYCHLY2018, Vignolo16}. For the successive approximates of 1D crystal or quasicrystal (i.e., taking larger unit cell), the IDOS is a step-like function where the steps become finer with the increasing size of the approximates. 
Constant frequency ranges in the ${\rm IDOS}(f)$ corresponds to the frequency gap of the system for $k=0$
The width of these ranges converges with larger approximates. The other feature allowing the identification of the frequency gaps is a specific character of ${\rm IDOS}(f)$ close to the gap's edges.
The changes of the frequencies for successive modes (i.e., with increasing IDOS)  become extremely small in the vicinity of the gap, which is the manifestation of van Hove singularities in density of states for 1D non-defected systems\cite{Hove53, Ashcroft76}. 
It is worth noting that, due to the lack of translational symmetry in quasicrystals, we cannot easily relate the frequency $f$ to the wavenumber $k$. However, it was shown that for 1D infinite system, the ${\rm IDOS}(f)\propto k(f)$ \cite{Lisiecki19a}. Therefore, the ${\rm IDOS}(f)$ dependence for large approximates give us the insight into the dispersion relation $f(k)$ -- see Fig.~\ref{fig:bragg}(a-c). 

The $f({\rm IDOS})$, i.e., inverse function of $\rm IDOS(F)$ for non-defected approximate (composed of 377 strips) is presented in Fig.~\ref{fig:bragg}(a). 
The solid black line in Fig.~\ref{fig:bragg}(a--c) shows the dispersion relation $f(k)$ for infinite uniform thin film\cite{Kalinikos_1986}. 
Please note the split of the x-axis between IDOS and wavenumber.
The film was assumed to have effective material parameters, which are the volume averages of the constituent material parameters of Co and Py. It is clearly seen that the $f({\rm IDOS})$ follows the dispersion
relation $f(k)$. 
The agreement is very good for long SWs, in the so-called metamaterial regime ($k\rightarrow 0$). In this case, SWs are not that sensitive to a specific configuration of strips. Significant differences are observed when frequency gaps are opened, which does not appear in the homogeneous film. Just before and after frequency gaps, differences between the frequencies of successive states are very small, and bars in the graph, Fig.~\ref{fig:bragg}(a), form the horizontal lines, which corresponds to the van Hove singularities. 

\begin{figure*}
\begin{centering}
\includegraphics[width=0.8\paperwidth]{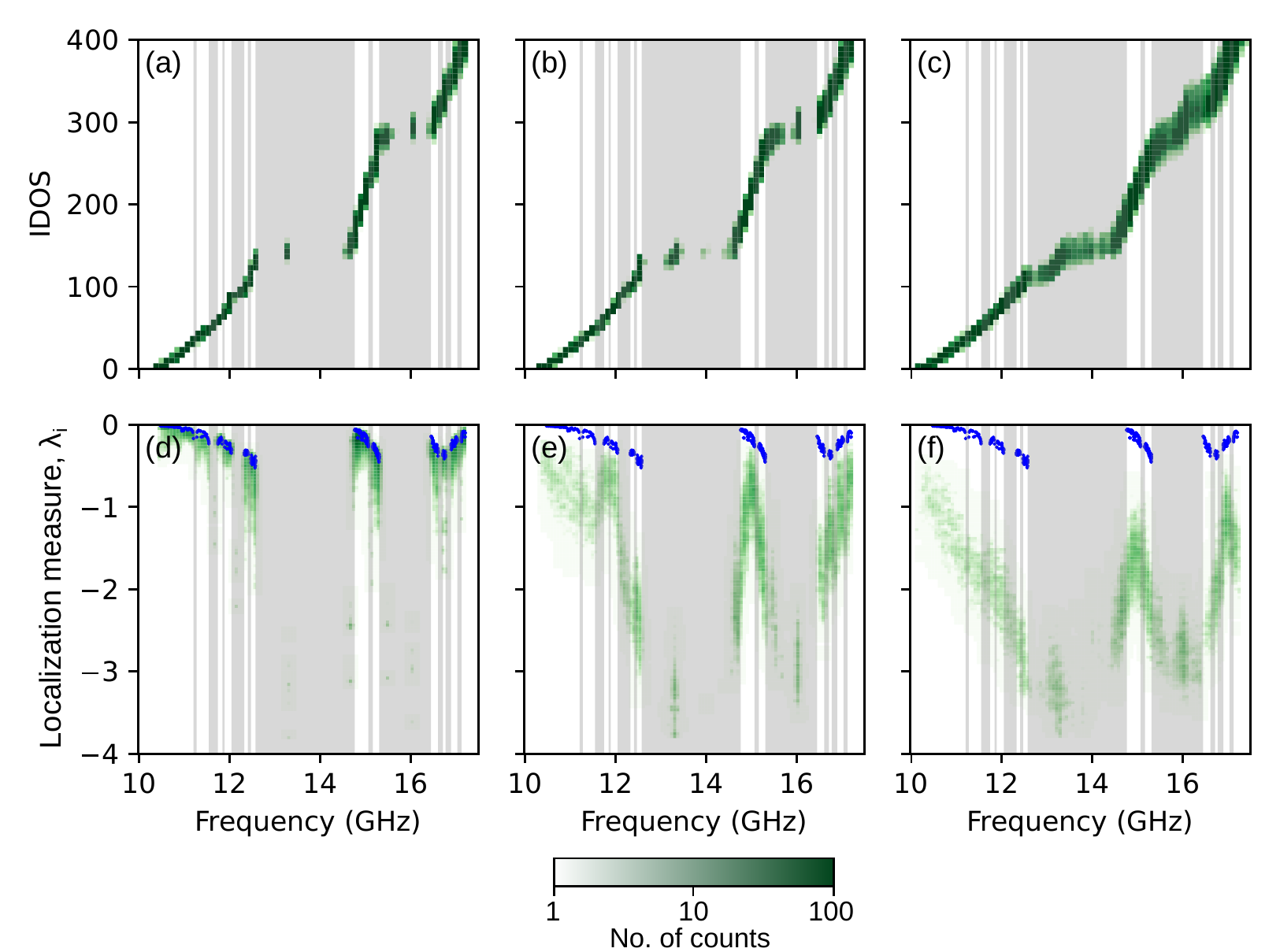}\caption{\label{fig:idos_loc}
Top row: (a), (b), and (c) -- integrated density of states as a function of frequency plotted in the inverse form $f(\rm IDOS)$ calculated
for Fibonacci sequence with introduced defects.
Gray areas represents the frequency gaps in ideal Fibonacci sequence.
The amplitude of phasonic defects $\Delta\phi/(2\pi)$ are: (a) $5\%$,
(b) $10\%$ and (c) $25\%$. 
Histogram of integrated density of states (IDOS) is obtained from 100 configurations of differently introduced defects. Intensity of green color reflects how often given position is occupied by SW mode.
Bottom row: (d), (e), and (f) -- localization measure $\lambda_i$  as a function of frequency for SW in 1D Fibonacci sequence with phasonic defects. The values of $\lambda_i$ are calculated for structures with (d) $5\%$, (e) $10\%$ and (f) $25\%$ of defects. Every plot aggregates $100$ different system configurations.
$\lambda_i$ increases significantly even if small amount of defects is introduced (d), and consequently
increases with grow amount of defects (e)-(f). 
}
\par\end{centering}
\end{figure*}

Figs.~\ref{fig:bragg}(b,c) show $f({\rm IDOS})$ in the presence of phasonic defects. We used green color for IDOS to visually differentiate results from non-defected case in Fig.~\ref{fig:bragg}(a). We consider two levels of phasonic defects corresponding to different ranges $\Delta \phi$ of the random component of the parameter $\phi$, which describes the structural degree of freedom (see, Eq.~(\ref{eq:chi}) and Fig.~\ref{fig:phasons_introduction}). We assume the values $\Delta \phi/(2\pi)=5\%$ (Fig.~\ref{fig:bragg}(b)) and $10\%$ (Fig.~\ref{fig:bragg}(c)).
Due to phasonic defects, the narrowest gaps are closed, and new modes strongly localized at defects (see discussion in the further part of the paper) are induced (see the red-dashed frames in Fig.~\ref{fig:bragg}(b,c), showing the states within the frequency gaps). The narrower gaps are much more susceptible to disappearing with increasing disorder.

In the bottom row of Fig.~\ref{fig:bragg}(d--f) we present Fourier spectra of the  structures considered in Fig.~\ref{fig:bragg}(a--c).
Formation of the frequency gaps can be attributed to the fulfillment of the Bragg condition, i.e., the position of the Bragg peak (multiplied by two) determines the position of frequency gaps\cite{Limonov2012}. However, their intensity does not necessarily determine the width of the frequency gap. We can see in the unperturbed Fibonacci structure (see Fig.~\ref{fig:bragg}(a) and (d)) that the biggest peak (except for a peak at $k=0$) is responsible for the widest frequency gap (12.3 GHz--14.3~GHz), however the second biggest peak opens only a small one, around 15~GHz.
We can see that the Bragg peaks are reduced as the level of phasonic defect increases. The relative reduction of the highest peaks (corresponding to wider gaps) is smaller than for lower peaks (corresponding to narrower gaps). 
Therefore, only the highest peaks in the Fourier spectrum are distinguishable, and the widest gaps remain opened for a large level of phasonic defects -- see the bottom part of Fig.~\ref{fig:bragg}(c) and the zoomed region, marked by the red dashed frame. 
Another effect of the phasonic defects in IDOS is the change of the slope of $f({\rm IDOS})$ at the edges of frequency gaps. This means that density of states is not singular anymore at these points.

Fig.~\ref{fig:bragg}(a--c) shows that the $f({\rm IDOS})$ is an useful function for description of the spectral properties of defected quasicrystals. However, the spectra presented in Fig.~\ref{fig:bragg}(b,c) are specific for given, randomly generated, set of phasonic defects. 
To obtain the representative picture, we need to collect the spectra for many configurations of phasonic defects generated for the same amplitude $\Delta \phi$. 
Figs.~\ref{fig:idos_loc}(a--c) present the IDOS for 100 different configurations aggregated on one plot in the form of 2D-histogram. 
Please note, that figures in two rows of Fig.~\ref{fig:idos_loc} share the same values of frequency on the horizontal axis.
The intensity of the green color reflects which position in frequency and IDOS appear more often. 
Figs.~\ref{fig:idos_loc}(a--c) are plotted for $\Delta \phi/(2\pi)=5,10,$ and $25\%$, respectively.
The gray background marks the frequency gaps of the non-defected Fibonacci sequence. 
The general trend of ${\rm IDOS}$ in the function of frequency prevails even for the most disturbed system. 
The IDOS curve is not much dispersed, suggesting the same spectra for the different realizations of the disorder.
However, we can notice that the green line in Fig. \ref{fig:idos_loc}(c) is thicker than in Fig. \ref{fig:idos_loc}(a), which indicates some frequency shift under strong disorder.
In the range of frequency 10--12 GHz, where IDOS resembles the dispersion relation of the homogeneous film with weight averaged material parameters (black line in Fig. \ref{fig:bragg}(a--c)), defects do not change the picture. 
The impact of the defects is strongest around the frequency gaps.
Initially, for $\Delta\phi/(2\pi) = 5\%$ the modes appear deeply inside and at the edges of the gaps. 
Then, for higher $\Delta \phi$, the modes start occupying other frequencies within the gaps and gradually fill them. These effects are more effective for narrower gaps.
Finally, we do not observe the fine structure of the gaps in the spectrum which was a hallmark of quasiperiodicity. 
The location of the defect in the sequence and its neighborhood determines the frequency of strongly localized defect modes. For $\Delta \phi/(2\pi) = 5\%$ (Fig. \ref{fig:idos_loc}(a)) modes from the widest frequency gaps 
(i.e., the gap around the 13 GHz or 16 GHz) are induced by those phasonic defects which form the sequence of double Py strips. Thus, their position on IDOS is very specific. Moreover, since such sequence of strips is common in defected sequence, the modes are highly degenerated.
For more distorted sequence presented in Fig. \ref{fig:idos_loc}(b) and (c), different sequences become available like triple Py strips, so defect states can occupy other frequencies. 

\begin{figure}[t]
\begin{centering}
\includegraphics[width=1\columnwidth]{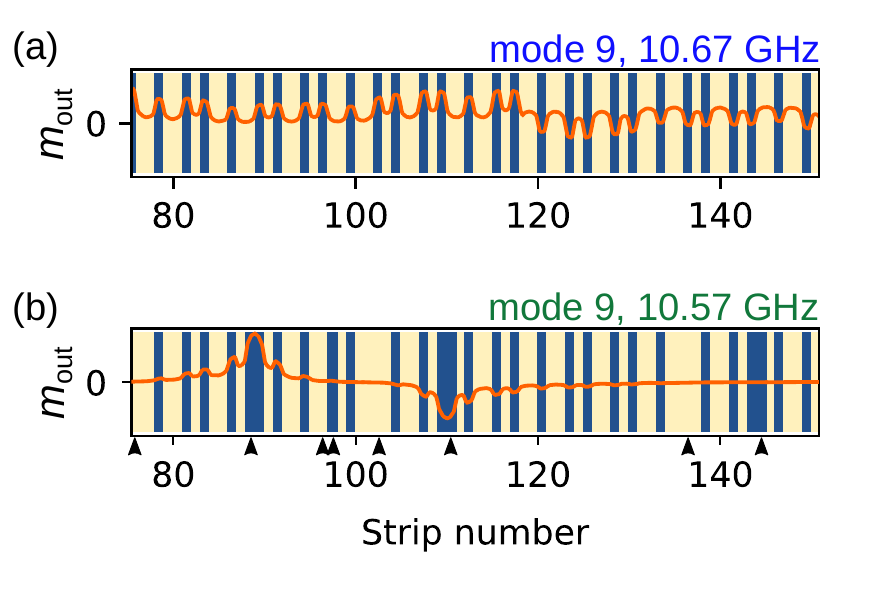}\caption{\label{fig: mod_set1}
The evolution of the bulk mode under the presence of the defects. (a)  In the absence of defects the mode is not localized, its amplitude  is more concentrated in Py than in Co. (b) For $\Delta \phi/(2\pi)=10\%$ the defects (marked by arrows below the plot) lead to the formation of double Py strips and can concentrate the SWs dynamics.
}
\par\end{centering}
\end{figure}

The qualitative determination of localization is challenging because the profiles of the SW modes can be localized in many regions, so the rate of spatial decay cannot be determined unambiguously. Therefore, we decided to introduce the global measure of localization $\lambda_{i}$, that is calculated for each ~$i^{\rm th}$ SW mode $\boldsymbol{m}_i(x)$:

\begin{equation}
\begin{split}
  \lambda_{i}&=-\frac{1}{L}\int_{0}^{L}\left|m_{i,\rm out}(x)\right|\log\left|m_{i,\rm out}\left(x\right)\right|dx,
\end{split}
\end{equation}
where $L$ denotes the width of the whole sequence. For the computational simplicity, we considered only the out-of-plane component  $m_{i,\rm out}(x)$ of dynamic part of magnetization $\boldsymbol{m}_i(x)$. During the calculations, the profiles are normalized: $\frac{1}{L}\int_{0}^{L}\left|m_{i,\rm out}(x)\right|dx=1$. 
The formulation of this measure is done with the analogy to the Shannon information entropy\cite{Mirbach95,Mirbach1998}, where SW profile plays a role of probability distribution -- the uniform distribution (and Dirac delta distribution) corresponds to the highest entropy and complete absence of localization: $\lambda_i$=0  (the lowest entropy and maximum localization: $\lambda_i=-\infty$).

In Figs.~\ref{fig:idos_loc}(d--f), we present the localization measure $\lambda_i$ for successive modes, calculated on the same data set as IDOS calculation.
They are ordered with increasing frequency, similarly to the IDOS spectrum. 
We can see that localization is significantly enhanced as the amplitude of phasonic defects is increasing (green 2D histogram in Fig.~\ref{fig:idos_loc}(d--f)), especially if we compare to the case of the non-defected system (blue points in Fig.~\ref{fig:idos_loc}(d--f)).
We can identify the strongly localized defect modes with a large value of $|\lambda_i|$ inside the frequency gap. 
It is worth noting that the localization of the modes at frequencies close to the edges of gaps with enhanced $\lambda_i$ suggests that some of the critically localized modes\cite{Kohmoto87,Macia96,DalNegro2003,Aynaou20} become defect modes. 

\begin{figure}[t]
\begin{centering}
\includegraphics[width=1\columnwidth]{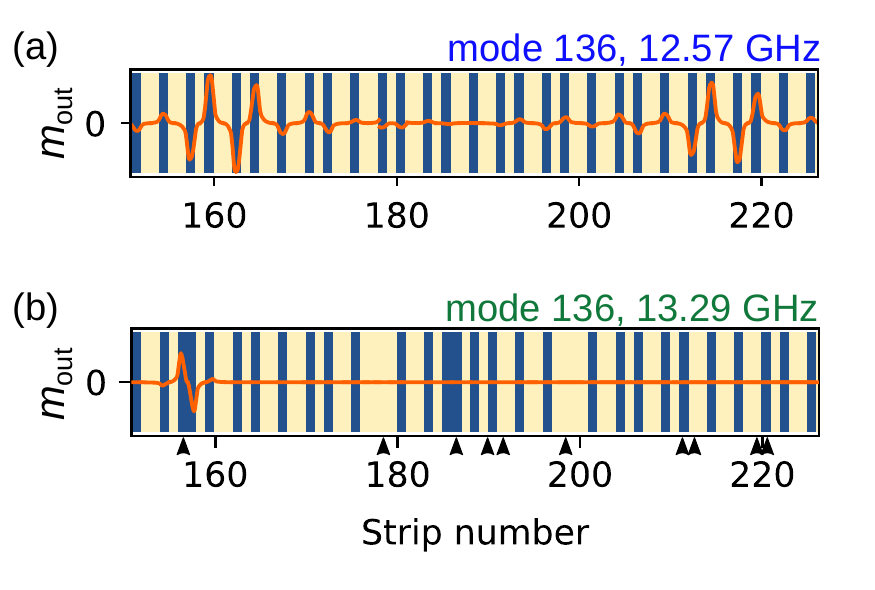}\caption{\label{fig: mod_set2}
The transition from the critical localization at the edge of the frequency gap to a strong localization in the frequency gap induced by phasonic defects. The critically localized mode (a) enters into the gap and become strongly localized (b), due to the presence of defects  $\Delta \phi/(2\pi)=10\%$.
}
\par\end{centering}
\end{figure}
To inspect the localization of the SW modes directly, we plotted the profiles of selected modes.
We chose one of the configurations for $\Delta \phi = 10\%$ that corresponds to intermediate disorder level, presented in Fig. \ref{fig:idos_loc}(b) and (e). 
All the modes are normalized to the maximum absolute value in whole structure. Figures presents only fragments of them, and the location can be deduced from strip numbers.
All  modes, which were selected for plotting, are also marked in the spectra (Fig. \ref{fig:bragg}(a,c))
We start the analysis by checking the impact of the disorder on the bulk modes. 
Fig.~\ref{fig: mod_set1} presents the comparison of one mode, labeled by No. 9 at 10.67 and 10.57 GHz in non-defected and defected structures, respectively. Looking at Fig. \ref{fig:idos_loc}(e) 
suggests significant modification of the profile. The envelope in Fig.~\ref{fig: mod_set1}(a) is not localized and the mode has several nodal points (one of them is visible close to the strip No. 120). The visible nonuniformity of amplitude is related to the oscillatory and evanescent behavior in Py and Co strips respectively, thus SW amplitude is concentrated in Py strips. 
Fig. \ref{fig: mod_set1}(b) presents mode No. 9 after introducing the defects, where a double Py strips are formed. 
SW is localized on the defects, around the strips No. 90 and 110 that have similar local arrangement. 

In non-defected Fibonacci quasicrystals, the critically localized modes exist close to the edges of the gaps --  see mode No. 136 at 12.57 GHz in Fig.~\ref{fig: mod_set2}(a) and its frequency marked in Fig.~\ref{fig:bragg}(c). The profile of this mode exhibits the pattern with amplitude concentrated on parts of the structure possessing locally the same arrangement of strips. For very large structures, these modes can reveal a self-similar pattern\cite{Kohmoto87,Aynaou20}.
By adding the defects, we can shift critically localized modes to the frequency gap. Then, their frequencies are changed significantly, and the profiles are extremely strong localized at defects -- see Fig.~\ref{fig: mod_set2}(b). The SW in Fig.~\ref{fig: mod_set2}(b) is localized in double Py, and since such defects occur several times within the considered structure, the mode can occupy different defects leading to multiple degenerations. 

\begin{figure}
\begin{centering}
\includegraphics[width=1\columnwidth]{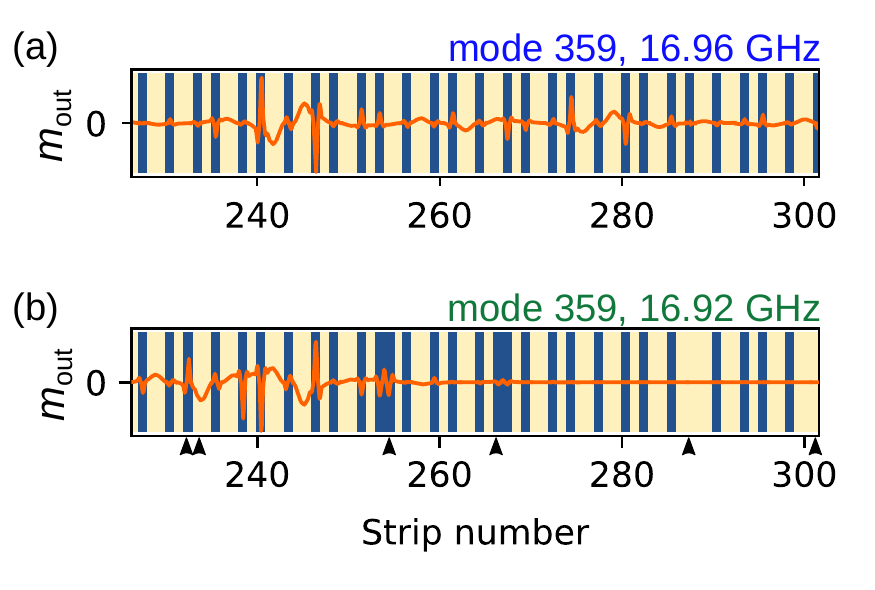}\caption{\label{fig: mod_set3}
(a) The critically localized mode (No. 359 at 16.96 GHz) which increases its localization due to partial confinement between defects (b) at $\Delta \phi/(2\pi)=10\%$.
}
\par\end{centering}
\end{figure}

The bulk modes can also increase their localization due to partial confinement between the defects. Fig.~\ref{fig: mod_set3}(a) presents the critically localized mode No. 359 at 16.96 GHz, which has enhanced amplitude on the sequences Co|Co|Py|Co (or on their reversed copies Co|Py|Co|Co). After introducing the defects, the mode amplitude is redistributed among these strips, which leads to the partial confinement of this mode between the defects -- see~Fig.~\ref{fig: mod_set3}(b). 

\begin{figure}
\begin{centering}
\includegraphics[width=1\columnwidth]{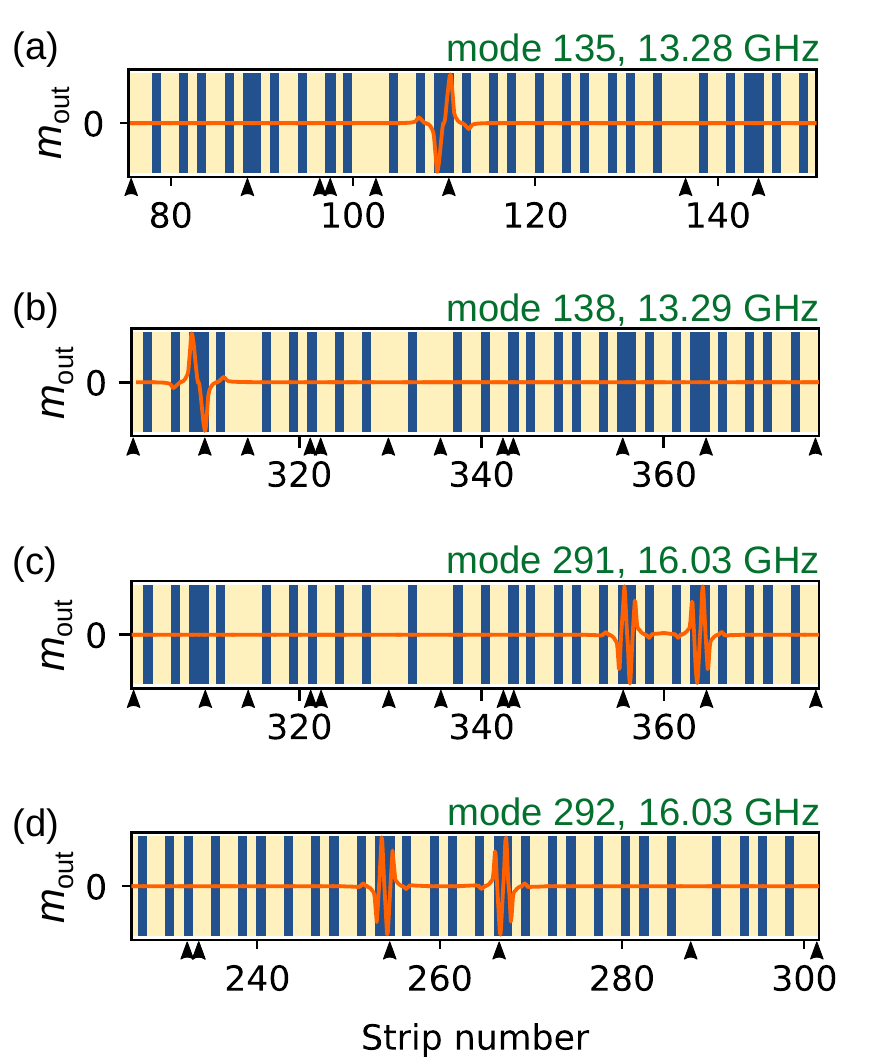}\caption{\label{fig: mod_set4}
The defect modes from the two largest frequency gaps shown in Figs.~\ref{fig:bragg} and \ref{fig:idos_loc}.
(a), and (b) modes No. 135 and 138 with frequency 13.3 GHz. (c), and (d) modes No. 291 and 292 with frequency 16 GHz.
The modes are strongly localized at one of few location but have the same profile, differing only in phase (flipped up-side-down) -- see (a,b) or reversed along with the structure (flipped left-right) -- see (c,d). Due to strong localization, the modes are practically degenerated. The results are shown for the structure with $\Delta \phi/(2\pi)=10\%$.
}
\par\end{centering}
\end{figure}

The most typical kind of localization, existing in both periodic and quasiperiodic structures, is an exponential localization on defects, which are observed within the frequency gaps.
We selected two wide gaps, around the frequency 13.5 GHz or 16 GHz -- Figs.~\ref{fig:bragg}, and \ref{fig:idos_loc}, to investigate the profiles of defect modes. The selected modes (shown in Fig.~\ref{fig: mod_set4}) are localized at the defects, which have the form of double Py strips. We arbitrary chose the modes with one phase flip inside the single defect (Fig.~\ref{fig: mod_set4}(a,b)), and three phase's flips inside the defect (Fig.~\ref{fig: mod_set4}(c,d)).
The defect modes are located at single or few positions in the structure. Due to strong localization and low probability of overlapping between the profiles concentrated at selected defects, the modes are degenerated -- there are many modes of very similar frequencies, occupying similar sequences in different locations of the quasicrystal. We discussed earlier the position-dependent susceptibility for inducing the defects, where we showed that some locations in the structure are very resistant or even completely robust to the introduction of defects at the low value of the amplitude $\Delta \phi$\cite{Naumis2007}. This is an additional factor supporting the isolation of the SW dynamics at defects and contributing to the non-uniform distribution of the frequencies for defect modes within the frequency gaps.

\section{Summary}
It is known that magnonic quasicrystals offer additional possibilities in designing artificial magnonic band structures as compared to magnonic crystals. The increased complexity of the spin-wave spectrum and the appearance of bulk localization of the spin-wave modes are the main effects of the quasiperiodicity. In the paper, we show additional steps towards customization, namely the introduction of the disorder in the form of phasonic defects, and demonstrate their impact on spectral properties and localization of the spin-wave modes. 
To explore the role of disorder in quasicrystals, we studied many randomly generated configurations of defects. We focused on selected configurations to discuss the profiles of representative eigenmodes exhibiting the critical localization at the edges of the frequency gaps, and strong localization on phasonic defects inside the gaps. 
In particular, we show that smaller gaps are closed under a small perturbation of the quasiperiodicity, while wide ones are relatively robust to a disorder. It is assisted by transition from bulk modes to critically localized modes, and finally to the modes strongly localized on the defects. Interestingly, the modes from the frequency gap edges become strongly localized by the introduction of phasonic defects to the structure, which is correlated with the disappearance of van Hove singularities. 

We demonstrated that in the complex magnonic system, where both short-range exchange interactions and long-range dipolar interactions come into play, the effects like closing the small gaps and enhancement of the modes' localization, are reproduced for spin-waves. The study opens the route for the investigation of phasonic defects in two-dimensional magnonic quasicrystals, which recently attracted interest due to their application potential in magnonics signal processing \cite{Watanabe20, Watanabe21}.

\section{acknowledgments}
S.M. and J.W.K. would like to thank Radosław Strzałka for fruitful discussion. All authors would like to acknowledge the financial support from the National Science Centre, Poland (projects:  No.~2020/36/T/ST3/00542,  No.~2020/37/B/ST3/03936, and No.~2020/39/O/ST5/02110). 

\appendix
\section{Cut-and-project method -- phasons\label{sec:app-C-P}}
\begin{figure}[b]
\begin{centering}
\includegraphics[width=1\columnwidth]{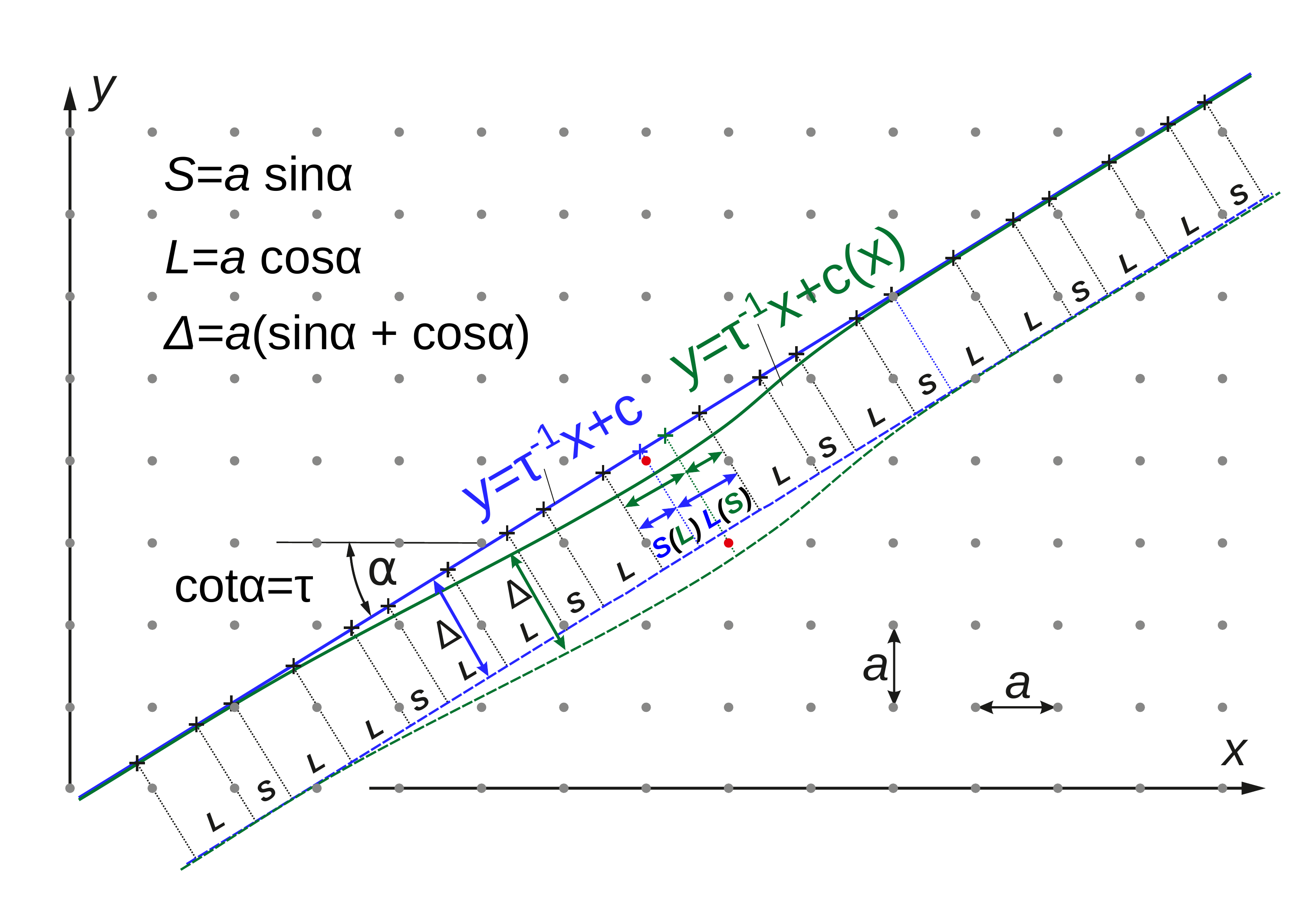}\caption{\label{fig: CaP}
The illustration of the cut-and-projection (C\&P) method and the induction of phasonic defect. The array of dots represent the square lattice in a 2D hyperspace. The Fibonacci lattice (black and blue crosses) is generated by the projection of square lattice from the belt between solid and dashed lines onto the line $y=\tau^{-1} x +c$ of irrational slope, being the inverse of the golden ratio $\tau$.  The visible (21-element) section of Fibonacci lattice corresponds to the selection of $\phi=2\pi c/a=0.8$, see Fig.~\ref{fig:phasons_introduction}.   For a defect-free Fibonacci lattice the belt (between solid blue and dashed blue line) is straight. By bending the belt (limited here by solid green and dashed green lines), we can induce the phasonic defects in the   Fibonacci lattice (black and green crosses).
}
\par\end{centering}
\end{figure}
The  Fibonacci lattice  can be generated from the square lattice of the period  $a$ by
  C\&P method\cite{Janot2012}. The lattice points $\mathbf{r}=a(m \hat{\mathbf{x}} + n \hat{\mathbf{y}})$, where $m,n$ are integers, are projected onto the line $y=\tau^{-1}x+c$ from the belt, below  this line, of the width $a(\cos\alpha+\sin\alpha)=a(\tau+1)/\sqrt{\tau+2}$, where $\alpha=\rm{arccot}(\tau)$ is the angle between the line and the $x$-direction, and $\tau$ is the golden ratio. This procedure generates the proper sequence of long  ($L=a\cos\alpha=a\tau/\sqrt{\tau+2}$) and short distances ($S=a\sin\alpha=a/\sqrt{\tau+2}$) between lattice points projected onto the line $y$,
  forming the Fibonacci lattice -- see Fig.~\ref{fig: CaP}. The position of the line (given by the constant $c$) and the related shift in the  perpendicular direction $\sqrt{\tau+2}(-\hat{\mathbf{x}}+\tau\hat{\mathbf{y}})$ express the structural degree of freedom in defining a Fibonacci lattice. Regardless on the value of this shift, we always obtain the defect-less lattices, differing only in some uniform translation of the lattice sites along the real (parallel) direction $\sqrt{\tau+2}(\tau\hat{\mathbf{x}}+\hat{\mathbf{y}})$. 
  
  The introduction of phasonic defect can be described by bending the belt. It is equivalent to the perturbation of structural degree of freedom, which can be expressed here as a position dependent shift of the belt: $c(x)$. When this dependence is small and smooth at the distances larger than the lattice constant $a$ then the phasonic defects have a form of the swaps between neighboring short and long distances in the Fibonacci lattice ($S\leftrightarrow L$).

\begin{figure}[t]
\begin{centering}
\includegraphics[width=0.9\columnwidth]{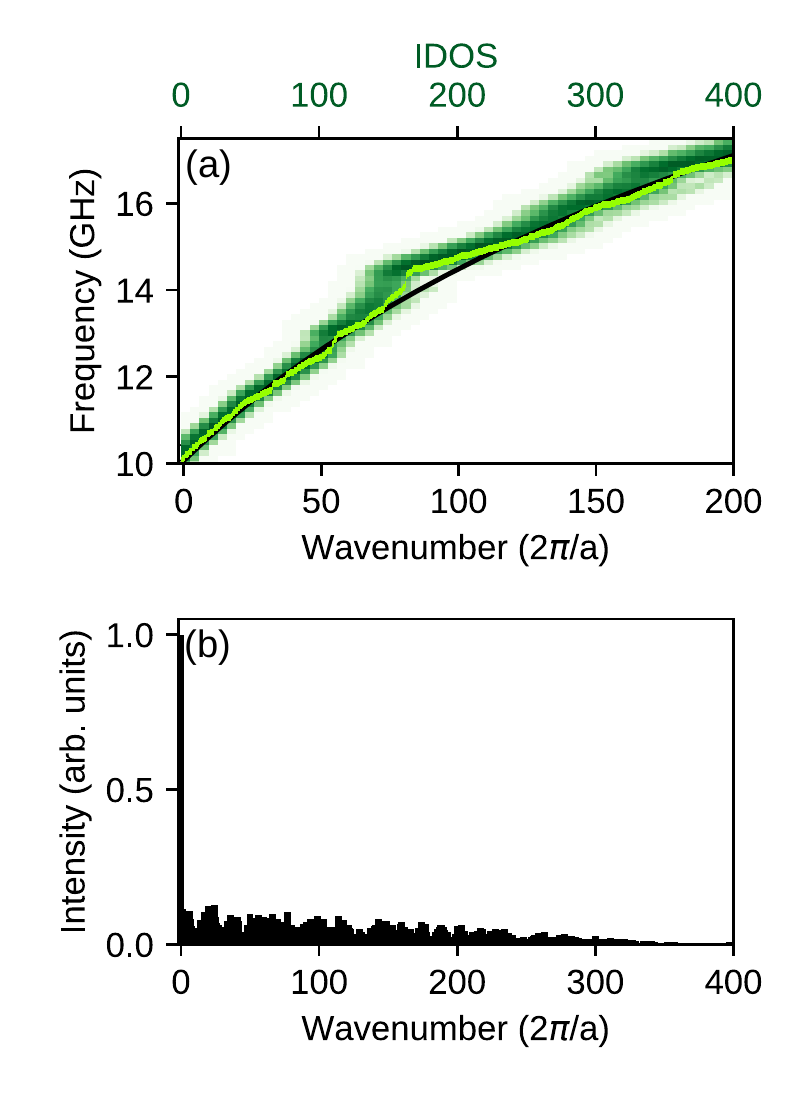}\caption{\label{fig:random}
(a) Integrated density of states (IDOS) for SWs in the randomly generated sequence of Co and Py, where the ratio between types of strips is kept as for Fibonacci quasicrystal, i.e., it corresponds to the golden ratio. The dark-green color represents a histogram of aggregated results obtained from 100 different random sequences (the color scale is the same as in Fig.~\ref{fig:idos_loc}).
Light-green points stand for one specific structure for which the bar of the Fourier transform (b) are plotted.
}
\par\end{centering}
\end{figure}

\section{Random system}

In Fig.~\ref{fig:random}(a), we show the IDOS spectrum of the SW eigenmodes in randomized sequence of Co and Py with the same parameter as in the paper. To keep the same averaged composition, We used 144 Py and 233 Co strips. 
We generated 100 different configurations, and intensity of green color reflect how often specific position is occupied on the plot.
By light green we plotted one selected configuration, for which Fourier spectrum is presented below.
The IDOS spectrum of this exemplary configuration coincides with the SW dispersion relation for uniform ferromagnetic layer with the volume averaged material parameters (i.e., with the weights $1/\tau$ and $1-1/\tau$), except small deviation around 14 GHz.
The IDOS does not show any signatures of the frequency gaps. It is also reflected in the Fourier spectrum (Fig.~\ref{fig:random}(b)) of this random structure that do not have any distinctive peaks except the peak at wavenumber $k=0$, which corresponds to the average value of the spatial distribution of material parameters. The absence of Bragg peaks is the signature of the lack of (quasi)crystal long-range order. 

The introduction of phasonic defect for large approximates of Fibonacci quasicrystal does not change the average number of Co and Py strips (it is obvious for the swaps Co$\leftrightarrow$Py whereas the substitutions Co$\rightarrow$Py and Py$\rightarrow$Co are equally probable -- see Fig.~\ref{fig:phasons_introduction}). In the limit $\Delta \phi\rightarrow 2\pi$ the IDOS spectrum approaches the spectrum of disordered system, as shown in Fig.~\ref{fig:random}(a).

\bibliography{bibliography}
\end{document}